# Convolutional Neural Network for Early Pulmonary Embolism Detection via Computed Tomography Pulmonary Angiography

Ching-Yuan Yu, Ming-Che Chang, Yun-Chien Cheng[#],
Department of Mechanical Engineering,
National Chiao-Tung University,
Hsinchu, Taiwan.

Chin Kuo[#],
Department of Oncology,
National Cheng Kung University Hospital,
College of Medicine, National Cheng Kung University,
Tainan, Taiwan.

[#]Corresponding author

*Abstract*—This study was conducted to develop a computer-aided detection (CAD) system for triaging patients with pulmonary embolism (PE). The purpose of the system was to reduce the death rate during the waiting period. Computed tomography pulmonary angiography (CTPA) is used for PE diagnosis. Because CTPA reports require a radiologist to review the case and suggest further management, this creates a waiting period during which patients may die. Our proposed CAD method was thus designed to triage patients with PE from those without PE. In contrast to related studies involving CAD systems that identify key PE lesion images to expedite PE diagnosis, our system comprises a novel classification-model ensemble for PE detection and a segmentation model for PE lesion labeling. The models were trained using data from National Cheng Kung University Hospital and open resources. The classification model yielded 0.73 for receiver operating characteristic curve (accuracy = 0.85), while the mean intersection over union was 0.689 for the segmentation model. The proposed CAD system can distinguish between patients with and without PE and automatically label PE lesions to expedite PE diagnosis.

*Contribution*—The proposed CAD system for triaging patients with PE explores the possibility of adopting a multiple convolutional neural network architecture.

I. INTRODUCTION

Pulmonary embolism (PE) refers to an obstruction of the pulmonary arteries and can be caused by blood clots, tumors, fat, air, and tissue. Given the extensive range of inducing factors and few specific clinical signs and symptoms, PE diagnosis can be delayed. Early diagnosis and management of PE are critical because they can decrease mortality. Clinically, computed tomography pulmonary angiography (CTPA) is performed for a definitive diagnosis of PE after a differential diagnosis has been made . Because a radiologist is required to analyze a CTPA report and suggest further management, this creates a waiting period that is dangerous for patients. A study in Europe found that approximately 34% of patients with PE died suddenly before therapy was commenced or showed any effect [1]. Thus, developing a solution that could alert physicians in real time may reduce or possibly eliminate the waiting period. Studies have investigated the application of computer-aided detection (CAD) models in CTPA image analysis and PE detection. For example, Bouma et al. [2] investigated different machine learning classifiers for selecting features based on vessel and lumen shape, voxel intensity, and texture shape and size. Using 3D lung vessel trees extracted from 2D PE images, Özkan et al. [3] was able to reduce the false-positive rate by employing median filtering to eliminate candidates with only one image or an unexpected shape. Other studies that have similarly reduced the false-positive rate include Park et al. [4], who employed an artificial neural network classifier, k-NN detection scores, and a 3D-grouping-based scoring method to select features based on the intensity, shape, and boundary; and Tajbakhsh et al. [5], who adopted a convolutional neural network (CNN) as the classifier and proposed a vessel-aligned multiplanar imaging method that employed two image channels to provide greater detail on regions within candidate images and thus improve classification. Yang et al. [6] proposed a novel two-step method comprising two CNN structures, wherein the first step distinguishes PE candidate images from an entire CTA scan via 3D object detection, and the second step implements ResNet-18 as the classifier to remove false positives.

The shared purpose of these models is that they have all been aimed at expediting PE diagnosis. Their primary function is to identify key images of lesions in patients with PE. While these models help radiologists by reducing the time that would otherwise be required to manually review many images, they cannot triage patients with PE from those without PE. Therefore, the present study was aimed at developing a two-step process that explicitly focuses on PE diagnosis during triage. We employed CNNs to develop a two-step CAD system consisting of the following two components: a classification model for PE detection and a semantic segmentation model for PE region labeling. The major contributions of our research are (1) the classification model can immediately obtain a preliminary impression of PE or non-PE from a CTPA image, thus facilitating patient management; and (2) the semantic segmentation model can accurately label lesion locations, thus reducing labor and time for manual tracing. Our method achieved a precision of 0.8537 and recall of 0.8487 in per-image evaluations, with a sensitivity of 0.82 and specificity of 0.9 obtained in per-patient evaluations.

## II. METHOD

This study was approved by the Institutional Review Board of National Cheng Kung University Hospital (NCKUH) and was granted an exemption and waiver of informed consent.

### A. Classification model: PE detection

- **Model architecture:** The architecture was designed to distinguish PE images with high sensitivity. When evaluating symptoms in a PE diagnosis, physicians consider both local and global information in CTPA images. As the first architecture, we thus selected a dilated residual network (DRN) [7] as the backbone of the classification model. The DRN spreads the receptive field to reduce the computational complexity, with dilated convolution employed to present more information and improve feature representation (**Figure 1**). The second architecture (**Figure 2**) was MixNet [8], which achieves a similar effect by using a mixed depth-wise convolutional mechanism. **Figure 3** shows how the mixed depth-wise convolutional layer broadens the receptive field by applying different kernel sizes to each feature channel. To achieve higher sensitivity, we applied a model ensemble to balance the strengths and weaknesses of these models. The model inputs were CTPA images and the model output was average predictions. Our multiple ensemble strategy (**Figure 4**) improved the per-image prediction sensitivity by 3% and specificity by 2%.

- **Per-patient classification:** Patients underwent CTPA examination and were given a preliminary diagnosis of PE or non-PE. When an image was predicted as a PE image, our system classified that patient as a PE patient; otherwise, they were classified a non-PE patient. We input all CTPA images into the classification model. The results showed that approaching high sensitivity increased the false-positive prediction of non-PE patients in the per-patient tests. Most false-positive predictions were from images of the entire lung region. To refine the classification and reduce the false-positive rate in non-PE patients, we trained another DRN with a modified PE100 dataset comprising only PE images from patients with PE and images of the entire lung region from non-PE patients.

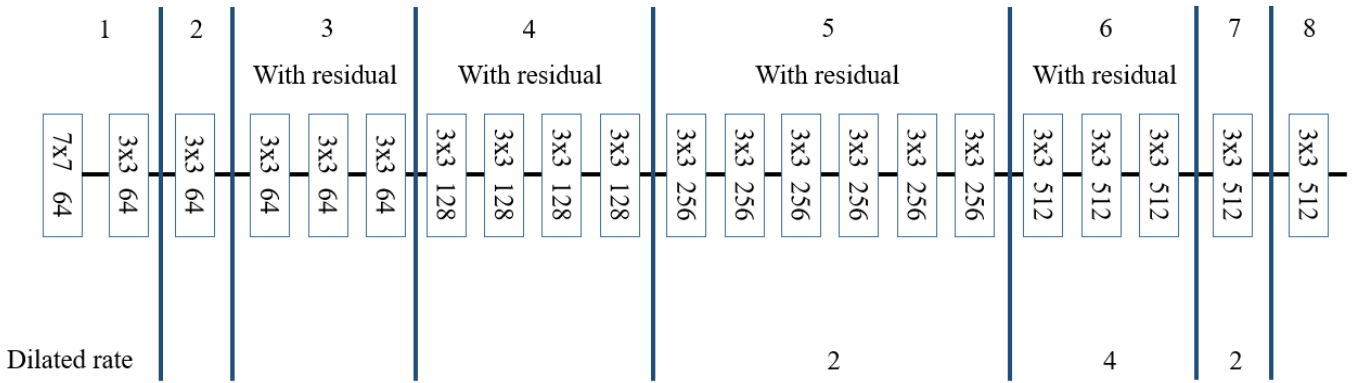

Figure 1: DRN architecture.

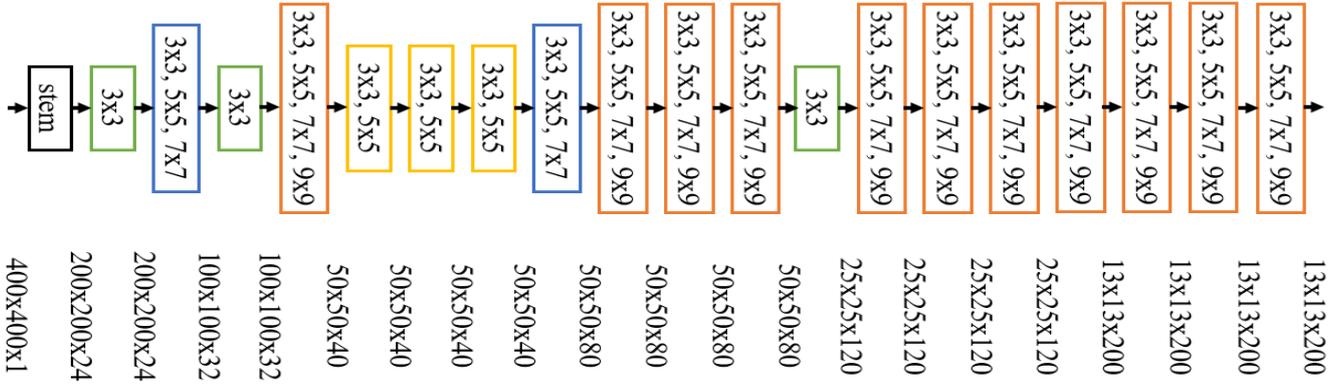

Figure 2: MixNet architecture.

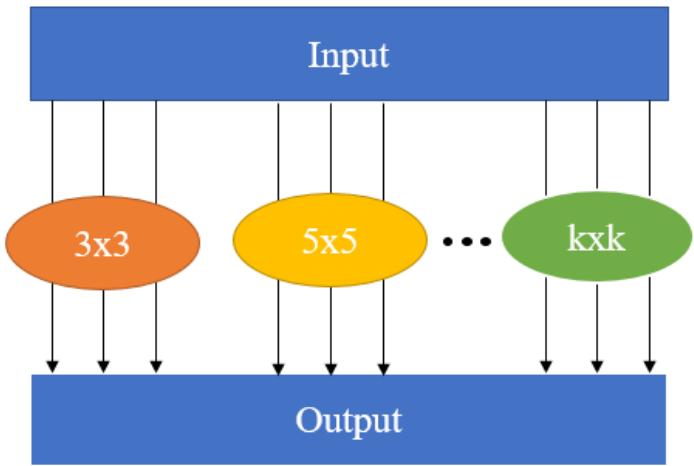

Figure 3: Mixed depth-wise convolutional layer.

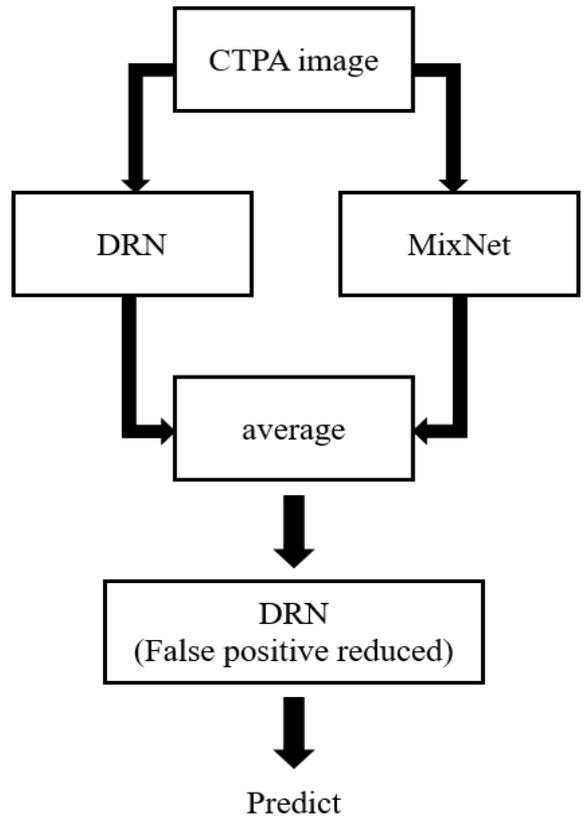

Figure 4: classification process.

B. *Semantic segmentation model: PE lesion pixel-wise segmentation*

- **Model architecture:** The semantic segmentation model was designed to label PE regions in images where PE was detected. PE typically occupies only a small region in an image, and this makes it difficult to label the region precisely. Thus, we prioritized acquiring precise location information over presenting complex high-level features. We adopted ARUX-Net, which is based on the ARU-Net [9] architecture, because it exhibits excellent performance in detecting minute historical text lines. ARU-Net [9] is based on U-Net [10]. It uses residual block and attention mechanism to solve the vanishing gradient problem, allowing the model to focus on appropriate spots. We used the squeeze-and-excitation layer [11] to enhance the feature extraction and applied joint pyramid upsampling (JPU) [12] for the feature decoding. As illustrated in **Figure 5**, ARUX-Net comprises two CNN structures, with RUX-Net as the encoder–decoder architecture and A-Net as the attention mechanism architecture. In A-Net, a four-layer basic CNN structure was adopted so that the model would focus on appropriate spots. We implemented an encoder–decoder architecture in RUX-Net using four downsampling layers, a JPU mechanism, and three upsampling layers. In each downsampling layer, two convolution sets and a squeeze-and-excitation layer were applied. Each convolution set comprised a 2D convolution layer, batch normalization layer, and activation layer (**Figure 4**). The squeeze-and-excitation layer is detailed in **Figure 5**, where the feature maps of Down3 and Down4 are concatenated, and the feature representation is refined via JPU. The feature map was then upsampled using three upsampling layers, and features extracted in Down1 and Down2 were combined using a shortcut in U-Net.

C. *Implementation details*

We coded and implemented the network using Python v3.6 with Tensorflow. The algorithm was trained on a system with two GTX 1080 Ti graphics processing units (GPUs) (12G per GPU) (Nvidia; Santa Clara, CA, USA). The method proposed by Ranger [13] was adopted to optimize the classification and segmentation models. The original learning rate was set at 1e-3, and the learning rate and momentum were automatically tuned in each epoch. Binary cross-entropy (BCE) with focal loss design [14] was selected as the loss function of the classification model, and BCE combined with dice loss [15] was adopted as the loss function of the segmentation model.

D. *Model evaluation and performance statistics*

Classification model performance was evaluated with the test set by using the area under the curve (AUC) of the receiver operating characteristic, precision, and recall per image. Precision and recall are defined in (1) and (2). Per patient test was evaluated with sensitivity, specificity, positive predictive value (PPV), and negative predictive value (NPV). The mean intersection-over-union (mean IoU) was employed to evaluate the semantic segmentation model performance by using the test set.

$$\frac{TP_0}{TP_0+FP_0} \times W_0 + \frac{TP_1}{TP_1+FP_1} \times W_1 \qquad (1)$$

$$\frac{TP_0}{TP_0+FN_0} \times W_0 + \frac{TP_1}{TP_1+FN_1} \times W_1 \qquad (2)$$

Here, $TP_1$, $TP_0$ denote true positive in the PE class and non-PE class; $FP_1, FP_0$ are false positive in the PE class and non-PE class; $FN_1, FN_0$ are false negative in the PE class and non-PE class; and $W_1, W_0$ are image amount ratios of the PE class and non-PE class.

III. EXPERIMENTS AND RESULTS

A. *Data and pre-processing*

We retrospectively reviewed consecutive patients who underwent CTPA examination. The dataset used to construct the PE detection model comprised data from NCKUH and an open resource [16]. The IRB number of NCKUH dataset is B-ER-108-380. The NCKUH dataset included patients who underwent a CTPA examination between January 1, 2018, and September 1, 2019. A senior radiologist and radiation oncologist reviewed and labeled the CTPA digital imaging and communications in medicine (DICOM) images. These physicians gave a diagnosis of PE according to the CTPA radiographic features. CTPA of PE will show a filling defect within the pulmonary vasculature. When the pulmonary artery is reviewed on its axial plane, the central filling defect from the thrombus appears surrounded by a thin rim of contrast or a thin stream of contrast adjacent to the embolus. The affected vessel might also be enlarged.

The dataset used to construct the classification model included 200 patients (100 with PE and 100 without PE), with 165 selected from the NCKUH dataset and 35 selected from the open source. We obtained 3892 images with PE and 20,211 images without PE and performed scans on Siemens CT scanners. **Table 1** presents the scanner specifications and image protocol details. CT scans were of diagnostic quality. They were obtained with the following settings: energy = 100 kVp, image thickness = 8 mm, pixel space = 0.49 mm, and Optiray contrast administered intravenously. The datasets were shuffled and divided by patient into three sets: training, validation, and testing (ratio = 7:2:1). For a similar number of patients, the number of non-PE images was more than four times the number of PE images. Thus, we upsampled the PE images five times to account for the data imbalance. We then center-cropped 400×400 regions in the CTPA images, adjusted the contrast by limiting the Hounsfield Unit values to 600, and linearly transformed all pixel values to [−1, 1].

The open-source dataset was employed to develop the PE lesion pixel-wise segmentation model. This set contained 2301 PE images. We applied random horizontal and vertical flipping and center-cropped 400×400 regions of the CTPA images for data augmentation. This dataset was divided by patient into the same three datasets: training, validation, and testing (ratio = 7:2:1).

B. *Results of classification model: PE detection*

The DRN [7] was trained to achieve a precision of 0.854, recall of 0.862, and AUC of 0.707. Using MixNet-L [8], we achieved a precision of 0.824, recall of 0.841, and AUC of 0.644. Applying the ensemble improved the classification model performance, achieving a precision of 0.858 and recall of 0.872. With the false-positive rate reduced, the performance improved, with a precision of 0.854, recall of 0.849, and AUC of 0.729 (**Table 2**). In the per-patient test, a sensitivity of 0.818 and specificity of 0.9 were achieved with the ensemble using the false-positive reduction architecture. Among 21 patients within the test dataset, 9 out 11 PE patients and 9 out of 10 non-PE patients were predicted correctly. One PE patient was erroneously predicted because of poor image quality, and the other PE patient was misdiagnosed due to a solitary small lesion. One non-PE case with prominent mediastinal lymph nodes was mistaken as PE. Overall, the per-patient classification performance was clinically acceptable. Furthermore, a PPV of 0.90 and NPV of 0.91 were achieved in the per-patient test. (Alert fatigue among clinicians less concerned with a high PPV.)

TABLE I. DATA CHARACTERISTICS

| Category | Data from NCKUH | Open source |
|---|---|---|
| Patient number (PE vs. non-PE) | 100 vs. 100 | 33 vs. 2 |
| Image number (PE vs. non-PE) | 1591 vs. 9828 | 2301 vs. 6491 |
| Manufacturer | Siemens | Neusoft |
| Slice thickness (cm) | 0.8 | 0.1-0.2 |
| Tube voltage (kVp) | 100 | 120 |
| Axial spatial resolution (pixels) | 512×642 | 512×512 |

TABLE II. PER-IMAGE TEST RESULT

| Model | Precision | Recall | AUC |
|---|---|---|---|
| DRN | 0.854 | 0.862 | 0.707 |
| MixNet | 0.824 | 0.841 | 0.644 |
| Ensemble | 0.858 | 0.872 | 0.687 |
| Ensemble with false-positive reduction | 0.854 | 0.849 | 0.729 |

*C. Semantic segmentation model: PE lesion pixel-wise segmentation*

Physicians can use semantic segmentation and employ our model to directly focus on the colored region shown in **Figure 5** in order to obtain a more efficient diagnosis. In the segmentation test set, we obtained 108 images with a mask, and the model achieved a mean IoU of 0.689.

*D. Clinical workflow incorporating the CAD system*

**Figure 6** illustrates the clinical workflow incorporating the proposed CAD system. Immediately after CTPA examination, CTPA DICOM images are input into the classification model for analysis. The multiple CNN-structure ensemble then analyzes each image to detect PE. The output is whether a patient has PE. Once a patient is found to have PE, their CTPA images are further analyzed using the semantic segmentation model, which labels PE regions in the detected images to reduce the effort of having to manually trace lesions. Thus, the proposed system facilitates early detection and treatment.

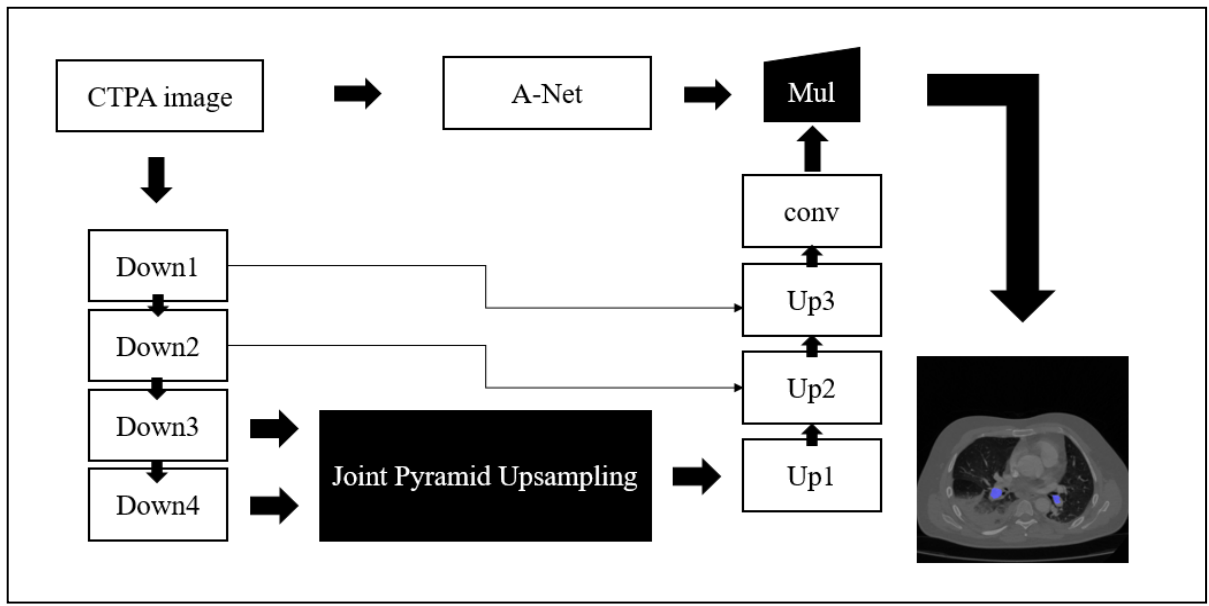

Figure 5. ARUX-Net architecture.

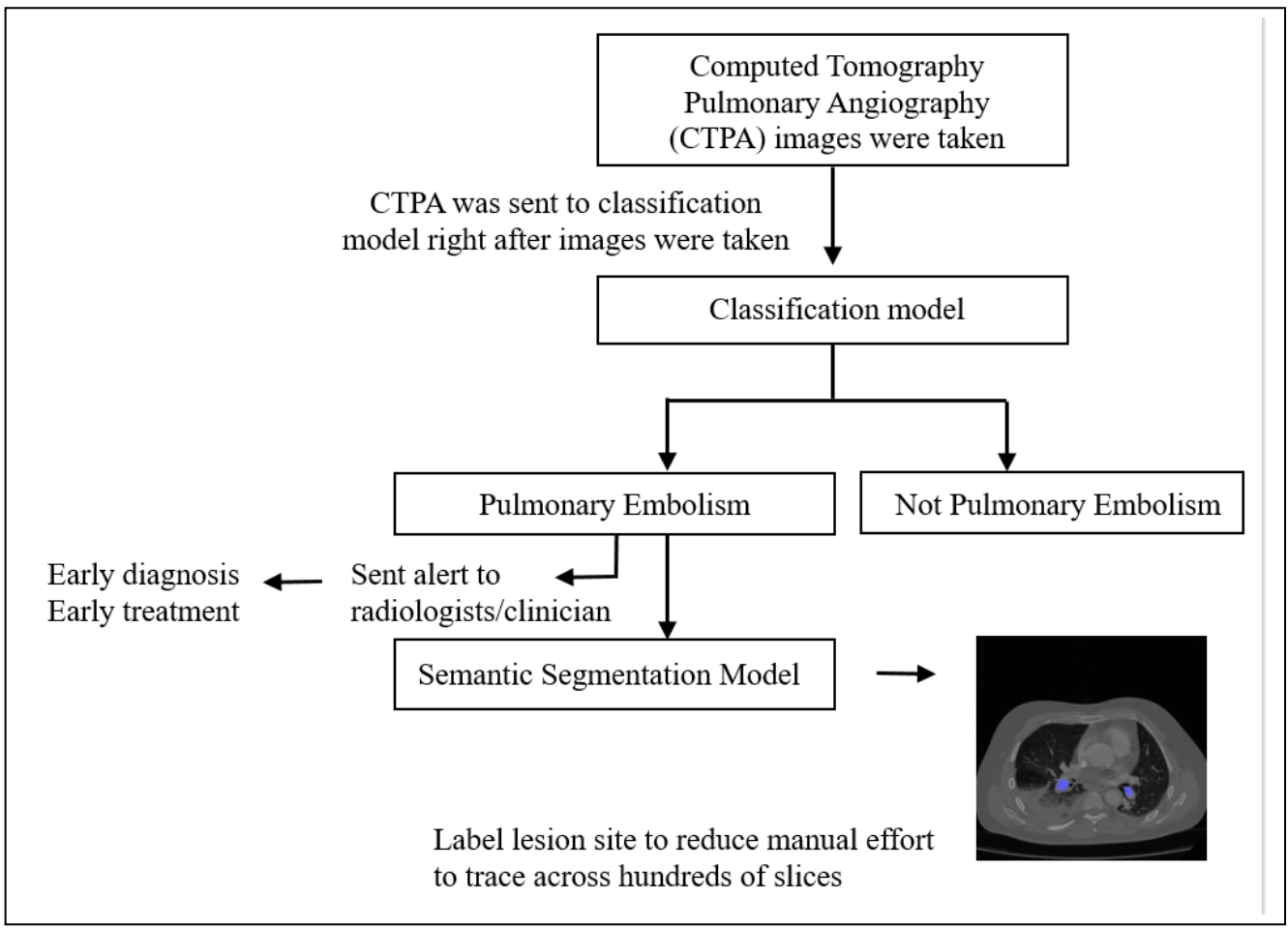

Figure 6. Proposed CAD system.

IV. CONCLUSIONS

This study tested a CAD triage system for the early detection of PE in patients who underwent CTPA examination. For the classification step, we proposed a model-ensemble architecture that achieved high precision, and employed a DRN to reduce the false-positive rate in non-PE patients. For the segmentation step, our model comprised multiple novel mechanisms that precisely label PE lesions with higher mIoU compared to what has been reported in previous studies. Our models successfully detected PE patients, labeled images with PE lesions, and fulfilled clinical demands related to early PE detection and treatment.

We plan to externally validate this model with another dataset in the future.

ACKNOWLEDGMENT

We express our gratitude for the kind assistance from the Department of Diagnostic Radiology, NCKUH, for offering the medical images and image labels.